\documentclass[11pt,twoside]{article}
\usepackage{asp2010}

\resetcounters

\begin{document}

\title{New Features in AST - a WCS Management and Manipulation Library}
\author{David S. Berry}
\author{Tim Jenness}
\affil{Joint Astronomy Centre, 660 N. A`oh\={o}k\={u} Place, HI, 96720, USA}

\begin{abstract}
Recent developments in the AST library are described, including a Python
interface, support for the FITS-WCS ``-TAB'' system for storing tabular
co-ordinate information, and extended support for representing
distortions in spatial projections, using several schemes in common use
(IRAF TNX/ZPX, Spitzer SIP, NOAO TPV and SCAMP).
\end{abstract}

\section{PyAST - a Python Wrapper for AST}

The Starlink AST library provides a comprehensive range of facilities for
attaching world co-ordinate systems to astronomical data, for retrieving
and interpreting that information in a variety of formats, including
FITS-WCS \citep{FITSWCS_1999}, and for generating graphical output based
on it. It is a mature system that has been presented at several ADASS
conferences over the past 14 years, ranging from \citet{AST_1998} to
\citet{AST_2009}. It is maintained by the Joint Astronomy Centre, Hawaii
(\url{www.jach.hawaii.edu}) as part of the Starlink Software Collection
\citep{SSC}. For further information about AST, see the AST homepage at
\url{www.starlink.ac.uk/ast}.

PyAST is a Python library that provides wrappers for the majority of
functions provided by AST. It is publicly available and can be downloaded
from github (\url{github.com/timj/starlink-pyasy/downloads}).
Documentation is available at \url{dsberry.github.com/starlink/pyast.html}.
PyAST requires Python 2.7 or later (version 3 is supported).

PyAST depends only on the numpy library (\url{numpy.scipy.org}). A copy
of AST is bundled with PyAST, so no other parts of the Starlink Software
Collection are required. Two optional interface are provided:

\begin{enumerate}

\item an interface for use with the AST Plot class that allows annotated axes
to be drawn using the popular matplotlib graphics library
\citep{MATPLOTLIB}. This has many
advantages over direct use of the axis annotation facilities provided by
matplotlib itself. For instance, axis labels can be placed within the
body of the plot, rather than round the edges (beneficial for many
projections, and essential for all-sky projections). Also, AST can draw
co-ordinate grids for projections that have peculiarities such as
singularities and discontinuities.

\item an interface for use with the AST FitsChan class that allows AST
to read and write FITS headers stored in a header object created by the
PyFITS library \citep{PYFITS_1999}.
\end{enumerate}

PyAST provides a few high level functions that wrap up other PyAST calls
to perform commonly required operations more easily. The HEALPix
grid \citep{HEALPIX}
shown in figure 1. could be produced with code similar to the following:

\begin{minipage}[b]{\linewidth}
\begin{verbatim}

>>> import pyfits
>>> import starlink.Atl as Atl
>>> import matplotlib.pyplot
>>>
>>> hdulist = pyfits.open( 'test.fit' )
>>> Atl.plotfitswcs( matplotlib.pyplot.figure().add_subplot(111),
>>>                  [ 0.1, 0.1, 0.9, 0.9 ], hdulist )
>>> matplotlib.pyplot.show()

\end{verbatim}
\end{minipage}

\articlefigure[scale=0.3]{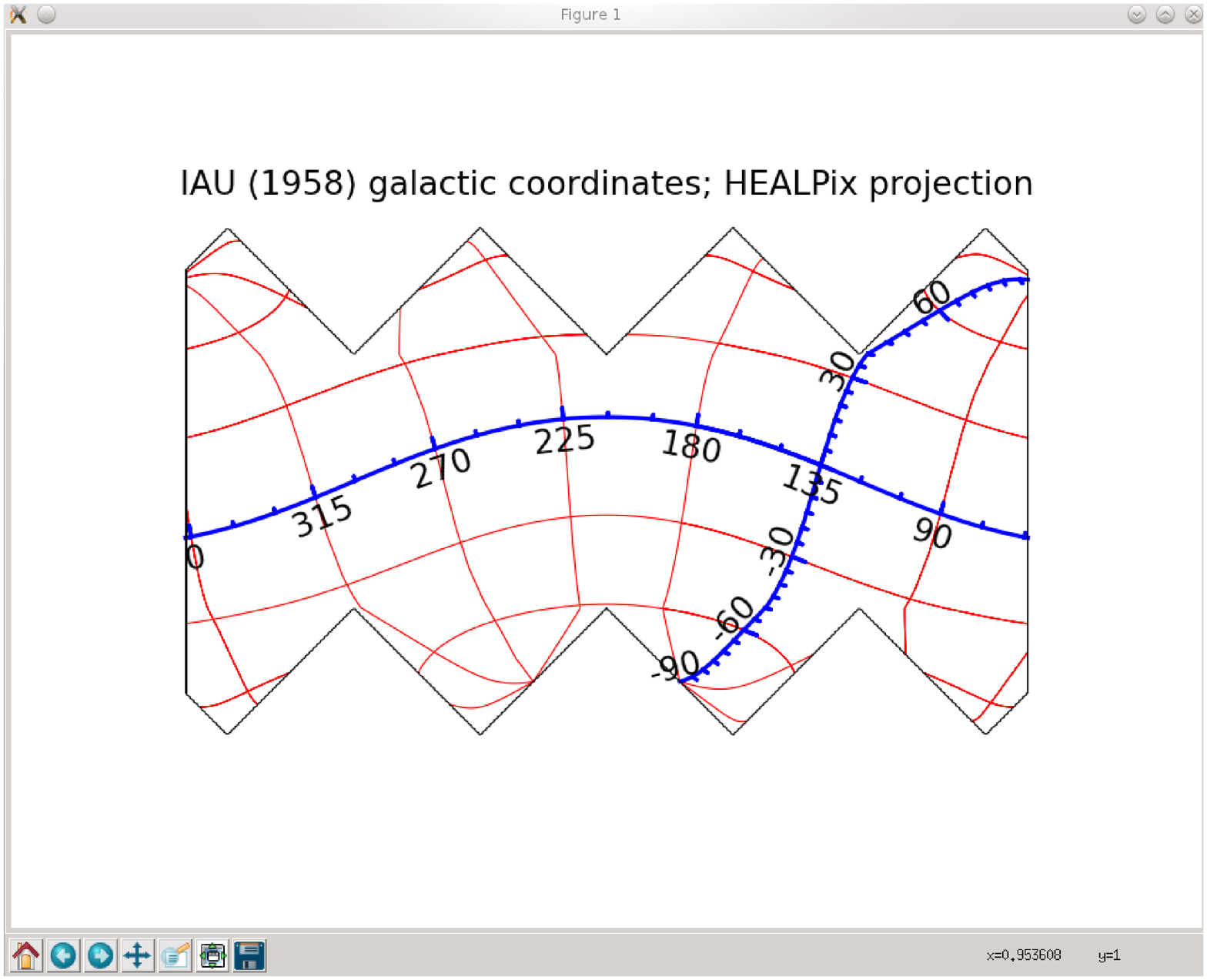}{}{PyAST displaying a HEALPix grid
in a matplotlib window}

\section{Support for Distorted Projections}
When published, FITS-WCS paper IV will address the issue of the
representation of distorted projections. But in the meantime, AST
supports several of the interim schemes that are in common use, as listed
below.

\begin{description}
\item[IRAF ``-TNX'':] AST can now reads TNX projections described by Chebyshev or simple polynomial with half-cross terms.
\item[IRAF ``-ZPX'':] AST can now reads ZPX projections described by Chebyshev or simple polynomial with half-cross terms.
\item[Spitzer ``-SIP'':] AST has been able to read SIP projections \citep{SIP} for
some time, but SIP support has been improved recently. Within a SIP
header, the forward and inverse transformations between world and pixel co-ordinates are defined by separate polynomials, but some SIP headers do not define an inverse transformation (from world to pixel co-ordinates). For such a header, AST can now implement an iterative inverse transformation.
\item[IRAF ``-TPV'':] AST now supports this renaming of the distorted TAN projection included in an early draft of FITS-WCS paper II.
\item[SCAMP ``-TAN'':] This is the same as the TPV projection, but
uses a CTYPE code of ``-TAN'' instead of ``-TPV''. AST differentiates
between SCAMP \citep{SCAMP} TAN headers and standard TAN headers by looking for PV
keywords attached to the latitude axis (a standard TAN projection should
have no such latitude PV keywords).
\item[AUTOASTROM ``-TAN'':] This is another representation of the TPV
projection, again using a CTYPE code of ``-TAN'', but using QV keywords instead of PV keywords to store the polynomial coefficients.
\end{description}

\section{DS9 and AST}
AST was written to provide co-ordinate handling facilities for the
Starlink software collection, including GAIA, SPLAT, KAPPA, \emph{etc.},
but it is now also used in other non-Starlink software. Particularly, the
DS9 image browser \citep{DS9} has for many years used AST to draw its
annotated co-ordinate grids. As of DS9 version 7.0 (currently in beta
testing), it will use AST additionally for all its WCS transformations,
thus benefiting from the improvements to support for distorted
projections listed above.

\section{Support for the FITS-WCS ``-TAB'' Algorithm}
AST now includes support for reading and writing tabular WCS information
in the form of FITS headers using the ``-TAB'' algorithm described in
FITS-WCS paper III. Currently, no support is included for the
multi-dimensional tables needed to describe non-separable axes.

\articlefiguretwo{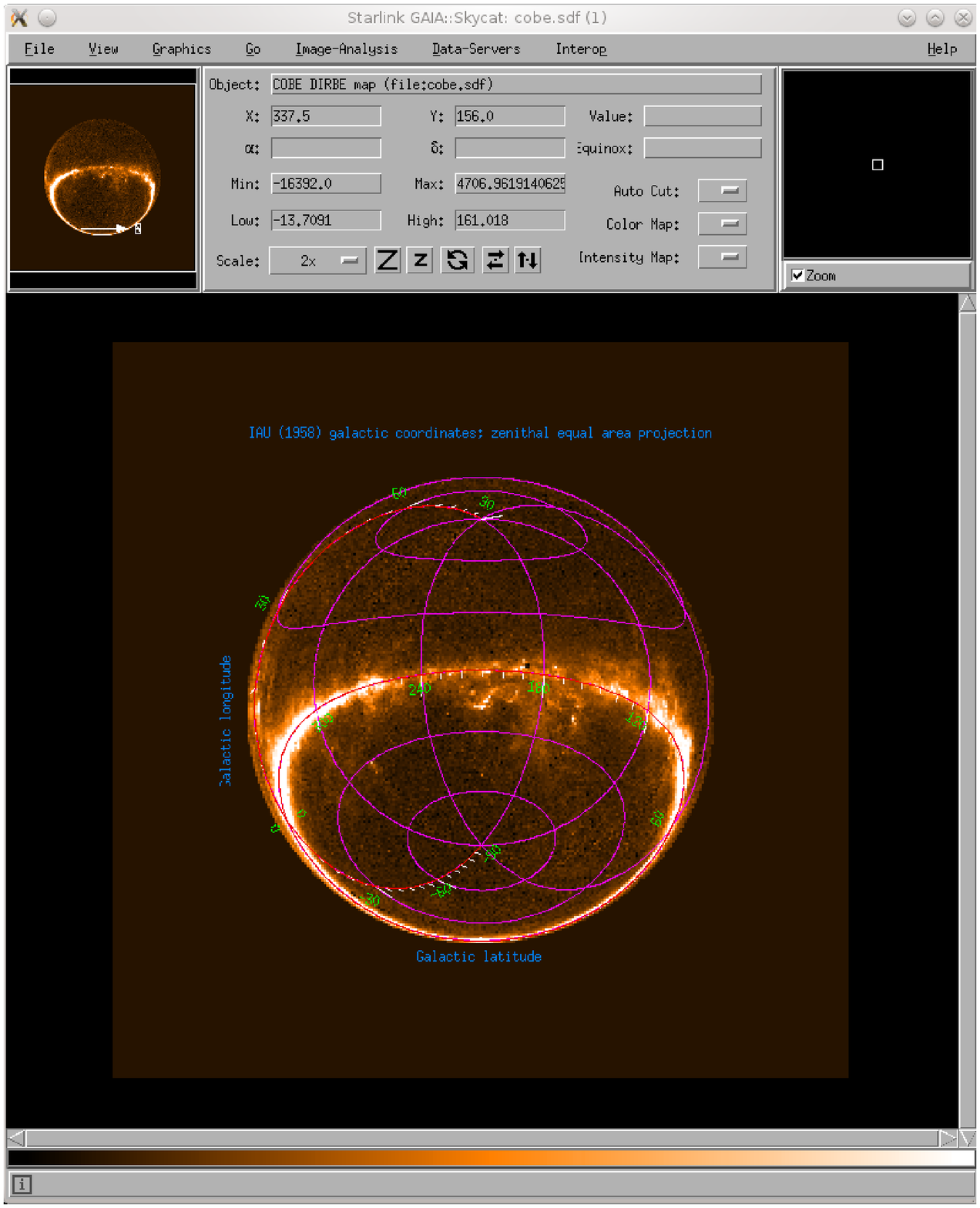}{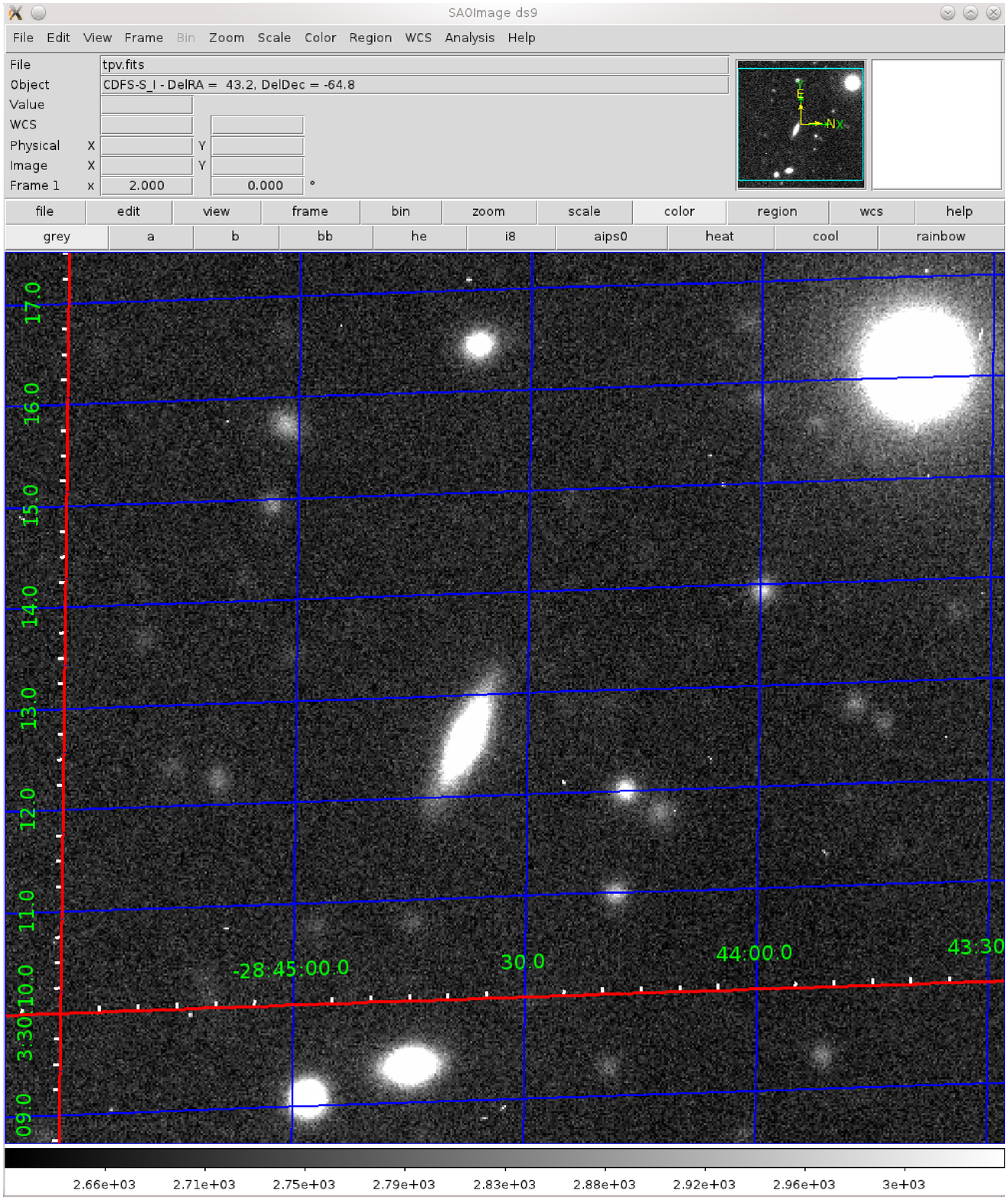}{}{Left: GAIA
using AST to display a co-ordinate grid for a ZEA (zenithal equal area)
projection. Right:DS9 using AST to display a co-ordinate grid for a
TPV (distorted TAN) projection.}

\bibliography{P011}

\end{document}